\documentclass[compsoc,journal]{IEEEtran}
\usepackage{cite}
\usepackage{amsmath,amssymb,amsfonts}
\usepackage{algorithmic}
\usepackage{graphicx}
\usepackage{textcomp}
\usepackage{seqsplit}
\usepackage{xcolor}
\usepackage{makecell}
\usepackage{enumitem}
\usepackage{babel}
\usepackage{hyperref}
\def\BibTeX{{\rm B\kern-.05em{\sc i\kern-.025em b}\kern-.08em
    T\kern-.1667em\lower.7ex\hbox{E}\kern-.125emX}}
    
\usepackage{amsthm}
\newtheoremstyle{cited}{3pt}{3pt}{\itshape}{}{\bfseries}{.}{.5em}{\thmname{#1} \thmnumber{#2}\thmnote{\normalfont#3}}
    
\newtheorem{theorem}{Theorem}
\theoremstyle{definition}
\newtheorem{definition}{Definition}
\newtheorem{conjecture}{Conjecture}
\newtheorem{problem}{Problem}
\theoremstyle{cited}
\newtheorem{citedthm}[theorem]{Theorem}

\newenvironment{hproof}{\proof}{\endproof}    

\makeatletter
\def\ps@IEEEtitlepagestyle{%
  \def\@oddfoot{\mycopyrightnotice}%
  \def\@evenfoot{}%
}
\def\mycopyrightnotice{%
  {\footnotesize This manuscript has been accepted for publication at IEEE Transactions on Computers DOI: 10.1109/TC.2021.3095669}
  \gdef\mycopyrightnotice{}
}

\begin{document}

\title{Fast Generation of RSA Keys using\\Smooth Integers}

\author{Vassil Dimitrov, Luigi Vigneri and Vidal Attias
\IEEEcompsocitemizethanks{\IEEEcompsocthanksitem V. Dimitrov is with IOTA Foundation and with the Department of Electrical and Computer Engineering at the University of Calgary (e-mail: vassil@iota.org, vdimitro@ucalgary.ca).
\IEEEcompsocthanksitem L. Vigneri and V. Attias are with IOTA Foundation (e-mail: first.last@iota.org).}
}

\IEEEtitleabstractindextext{
    \begin{abstract}
    Primality generation is the cornerstone of several essential cryptographic systems. The problem has been a subject of deep investigations, but there is still a substantial room for improvements. Typically, the algorithms used have two parts – trial divisions aimed at eliminating numbers with small prime factors and primality tests based on an easy-to-compute statement that is valid for primes and invalid for composites. In this paper, we will showcase a technique that will eliminate the first phase of the primality testing algorithms. The computational simulations show a reduction of the primality generation time by about 30\% in the case of 1024-bit RSA key pairs. This can be particularly beneficial in the case of decentralized environments for shared RSA keys as the initial trial division part of the key generation algorithms can be avoided at no cost. This also significantly reduces the communication complexity. Another essential contribution of the paper is the introduction of a new one-way function that is computationally simpler than the existing ones used in public-key cryptography. This function can be used to create new random number generators, and it also could be potentially used for designing entirely new public-key encryption systems.
    \end{abstract}
    
    \begin{IEEEkeywords}
    Multiple-base Representations, Public-Key Cryptography, Primality Testing, Computational Number Theory, RSA
    \end{IEEEkeywords}}

\maketitle

\section{Introduction}
 \IEEEPARstart{A}{dditive} number theory is a fascinating area of mathematics. In it one can find problems with extreme difficulties that can be posed in a relatively simple way. Goldbach or twin-primes conjectures are, perhaps, the simplest examples. One of the most challenging problems in modern number theory is the $abc$-conjecture, which can be posed as: the equation $a+b=c$ where $GCD(a,b) = 1$ does not have large solutions in numbers with only small prime factors. Benne de Weger’s thesis \cite{B.M.M.deWeger1989} provides a large number of fascinating specific numerical facts that highlight the properties of numbers with small prime factors only. For example, the equation $x+y=z$ has exactly 545 solutions in numbers of the form $2^a\cdot3^b\cdot5^c\cdot7^d\cdot11^e\cdot13^f$ such that $GCD(x,y)=1$ -- the largest one being $2^1\cdot3^{11}\cdot5^1 + 7^1\cdot13^1 = 11^6$. Numbers without large prime factors are usually called \textit{smooth numbers}. They have substantial use in various cryptographic primitives~\cite{bernstein2007, Dimitrov2007a, Dimitrov2011, Chabrier2013} as a tool to speed up specific operations (e.g., point multiplications over various types of elliptic curves) and also as a cryptanalytic tool (e.g., in the implementation of several algorithms for factoring and discrete logarithm problem).\newline

In this paper we will outline some properties of the smooth numbers and showcase how they can be used to improve the speed of various primality generation algorithms (see Section~\ref{sec:fast-gen-intro}), and, based on that, propose a one-way function which may be used to build new public-key encryption schemes (see Section~\ref{sec:one-way-intro}).

\subsection{Fast generation of prime numbers}\label{sec:fast-gen-intro}

The generation of prime numbers is a cornerstone of cryptographic systems such as the RSA cryptosystem. Although this problem has been deeply researched in the past~\cite{Rabin1980}, in this paper we show that further optimizations are possible, which may be of great interest especially in decentralized environments.\newline

Typically, the algorithms for primality generation have two parts: (i) trial divisions aimed at eliminating numbers with small prime factors, and (ii) primality tests based on an easy-to-compute statement that is valid for primes and invalid for composites. In this paper, we exploit the properties of smooth numbers to generate large random numbers that are void of small prime factors up to a particular limit. For example, if we are interested in generating 1024-bit primes, our algorithm can quickly produce a random number that is not divisible by the first, say, 100 primes. The existing primality testing algorithms usually implement trial divisions by small potential prime divisors of the number-to-be-tested before applying more powerful tests (e.g., Rabin-Miller~\cite{Rabin1980} or Solovay-Strassen~\cite{Solovay1977} tests). The point of the trial divisions is clear as there is no need to use expensive modular exponentiation-based primality testing for numbers that are obviously divisible by, say, 3 or 5.\newline

What is the maximum number of trial divisions of the algorithm one should perform? Surprisingly, it seems that this simple question has not been yet addressed by the computational number theory community. If we apply the trial division part for the first, say, 10,000 primes, and the number passes this first test, it has a higher chance to be a prime in comparison to a randomly selected odd integer. But the trial division part of such an algorithm will require a considerable amount of time. Instead of determining the optimal upper bound for the number of tested small prime divisors, we offer an algorithm that generates large random numbers that are not divisible by small primes up to a given limit. The algorithm can be used in both a centralized and a decentralized environment. The decentralized case is important for optimizing the implementation of Boneh-Franklin algorithm for efficient generation of shared RSA keys. This is extremely critical in many of the existing current blockchain architectures~\cite{cao2019strong,chen2019secure}.

\subsection{Alternative usages of smooth numbers}\label{sec:one-way-intro}

Whilst the usage of smooth numbers to speed up the primality generation provides a concrete and usable application, with this paper we aim to open areas for other potential uses of the smooth numbers in cryptographic applications. For instance, we conjecture that smooth numbers can be used to build a very simple one-way function. This new one-way function can be viewed as a dual computational problem to the factoring problem. One application is mathematically guaranteed to exist. A very powerful theorem by Hastad et al.~\cite{haastad1999pseudorandom} rigorously demonstrates that any one-way function can be used to build an efficient pseudo-random number generator. Exactly how this has to be done for our one-way function remains to be researched. Even more interesting is to investigate if one can obtain a public-key encryption scheme based on this function in a way, similar to the RSA algorithm.
 
\subsection{Outline} 

Our paper is structured as follows:

\begin{itemize}
\item Section~\ref{sec:smooth-numbers} discusses the main properties of the smooth numbers – density, difference between smooth numbers, and minimal representations.
\item Section~\ref{sec:one-way} introduces the new one-way function and compares the complexity of the associated forward (easy) and backward (hard) computational problems. 
\item Sections~\ref{sec:rsa-keys} and~\ref{sec:algo-improvements} are dedicated the new primality generation algorithm, based on the use of smooth integers. 
\item Section~\ref{sec:practical-considerations} describes practical considerations when our algorithm is implemented in specific applications.
\item Sections~\ref{sec:open-problems} and~\ref{sec:conclusions} discuss open problems and conclude the paper.
\end{itemize}

\section{Smooth numbers}\label{sec:smooth-numbers}

\subsection{Definitions and properties}

We introduce the following definitions:

\begin{definition}[$s$-integers]
    An integer number is called an $s$-integer if its largest prime factor is smaller than or equal to the $s$-th prime.
\end{definition}

\begin{definition}[Smooth integers]
    Numbers with only small prime factors, i.e., $s$-integers with $s$ small, are called \emph{smooth integers}.
\end{definition}

Before going into more general considerations, we show examples and properties of $1$- and $2$-integer numbers:
\begin{itemize}
    \item The perfect powers of 2 are 1-integers (their largest prime factor is $2$).
    \item Representation of an integer as the sum of different 1-integers is simply their binary representation. The representation is unique.
    \item The numbers from the sequence
        1, 2, 3, 4, 6, 8, 9, 12, 16, 18, 24, 27, 32, 36, 48, 54, 64, 72, 81, 96, \ldots,
    are 2-integers, since their largest prime factors is at most 3 (the second prime).
\end{itemize}

If one wants to represent an integer number as the sum of 2-integers, there are many possible representations. For example, 100 has \emph{exactly} 402 different representations as the sum of 2-integers. The shortest ones are $100 = 96 + 4 = 64 + 36$. The number of different representations of a given integer as the sum of numbers of the form $2^a\cdot 3^b$ (that is, 2-integers) can be predicted with extreme accuracy. For example, 40,000 has \emph{exactly} 2,611,771,518,060,603 different representations. This is proven from the following formula for $p(n)$, where $p(n)$ defines the number of different representations of $n$, and as the sum of 2-integers:

$$
p(n) = \left\{
    \begin{array}{ll}
        p(n-1)+p(n/3) & \mbox{if } n \equiv 0(\textrm{mod} \ 3), \\
        p(n-1) & \mbox{otherwise}.
    \end{array}
\right.
$$

This particular recursive equation was investigated by~\cite{Mahler1940} and an extremely accurate approximation of $p(n)$ was obtained by~\cite{Pennington1953}. A less accurate approximation of $\log p(n)$ is given by
\begin{equation}
\log p(n) \approx \frac{\log^2n}{2\cdot\log 3}.
\end{equation}

The most interesting representations are the sparsest ones (in a previous example we pointed out that for number 100, the sparsest representations are $96+4$ and $64+36$, requiring only two terms). The following theorem provides good information about the sparsity of the multiple-base representations:

\begin{citedthm}[\cite{Krenn2018}]
\label{multibaserepr}
Let $n$ be a positive integer. Then, it can be represented as the sum of at most $\mathcal{O}\left(\frac{\log n}{\log\log n}\right)$ 2-integers.
\end{citedthm}

Some facts about the sparsity of the representation of integers as the sum of 2-integers:

\begin{itemize}
\item23 is the smallest integer that cannot be represented as the sum of two 2-integers;
\item431 is the smallest integer that cannot be represented as the sum of three 2-integers;
\item18,431 is the smallest integer that cannot be represented as the sum of four 2-integers;
\item3,448,733 is the smallest integer that cannot be represented as the sum of five 2-integers
\item1,441,896,119 is the smallest integer that cannot be represented as the sum of six 2-integers.
\end{itemize}

In other words, any 30-bit integer can be represented as the sum of at most six numbers of the form $2^a3^b$, where $a$, $b$ are non-negative integers.
 
\subsection{The density of smooth integers}
The number of 2-integers less than $x$ is approximately:
\begin{equation}\label{eq:density-2int}
    \frac{\ln^2 x}{2\ln 2 \ln 3}.
\end{equation}
The number of 3-integers less than $x$ is approximately:
\begin{equation}
    \frac{\ln^3 x}{6 \ln 2 \ln 3 \ln 5}.
\end{equation}
The number of 4-integers less than $x$ is approximately:
\begin{equation}
    \frac{\ln^4 x}{24 \ln 2 \ln 3 \ln 5\ln 7}.
\end{equation}

Trivially, by extending the size of $s$, we can reduce the number of terms necessary to represent integers as the sum of $s$-integers. Many of the main theoretical properties of the smooth integers stem from the transcendental number theory and the theory of linear forms of logarithms. We will present some of the most essential ones here:

\begin{citedthm}[\cite{Tijdeman1974}]\label{th:tijdeman74}
Let $x$ and $y$ be two consecutive $s$-integers ($x>y$). Then their differences is bounded from above and below:
\begin{equation}
    \frac{x}{\log^{c_1}x} < x-y < \frac{x}{\log^{c_2}x},
\end{equation}
where $c_1$ and $c_2$ are effectively computable constants.
\end{citedthm}

Many results can be proved from this theorem. For example, one can prove that the multiplication by an $n$-bit constant can be achieved by using a sub-linear number of additions only, i.e., $\frac{n}{\log n}$.\newline

The following theorem provides information about the representation of integers as the sum of $s$-integers (see Appendix for an example):

\begin{citedthm}[\cite{Krenn2018}]\label{th:representation}
    Every integer $n$ can be represented as the sum of, at most, $\mathcal{O}\left(\frac{\log n}{\log\log n}\right)$ $s$-integers.
\end{citedthm}
\begin{hproof}
    Consider the case $s=2$, that is representing $n$ as numbers of the form $2^a\cdot 3^b$. According to Eq.~\eqref{eq:density-2int}, we know that the number of 2-integers in the interval $[2^{k-1}, 2^k]$ is approximately $0.63 \cdot k$. Theorem~\ref{th:tijdeman74} guarantees that they cannot be concentrated in a cluster. Therefore, if one applied a greedy algorithm to find a suitable representation, after subtracting $n$ minus the closest 2-integer,  we will get a number with $\mathcal{O}(\log\log n)$ bits less in its binary representation. Repeating the same procedure, one gets the bound in the above theorem. Please note that greedy algorithms in this case do not guarantee minimization, but the representations obtained by them is nevertheless asymptotically optimal.
\end{hproof}

The theorem, however, does not tell us anything about the constant associated with big-$\mathcal{O}$ notation. Experimental and probabilistic evidence suggests that it is probably equal to $2/s$ (see Table~\ref{tab:operations_cost}). This means that if $s$ is relatively large, then we might anticipate the representation of integers as the sum of very few smooth integers. When is \emph{very few} expected to become 2? That is, \emph{under what condition an integer $n$ can be written as the sum of two smooth integers?}

\begin{conjecture}
Let $n$ be an integer. Then, there exists a pair of integers $a$ and $b$ such that $a+b = n$ and their largest prime factor is $\mathcal{O}((\log n)^{2+\epsilon})$, where $\epsilon$ is any positive number.
\end{conjecture}

\textbf{We call this conjecture anti-Goldbach}. The original Goldbach conjecture states that every odd integer can be represented as the sum of three primes (proved for every sufficiently large odd number in 1937 by Vinogradov and unconditionally by~\cite{Helfgott}). Also, every even integer is the sum of two primes (still unproven). Prime numbers have the largest possible prime factors (themselves), whereas smooth numbers have only small prime factors, thus, the name of the conjecture. 

\subsection{Sum of smooth numbers}

The problem of representing numbers as sums of smooth integers was firstly considered by Erd\"os and Graham in\cite{erdos1980}. It is very interesting to point out that in the case of representing every sufficiently large integer as the sum of \textit{three} smooth numbers, sharper bounds for the smoothness of the summands are known, as opposed to the case of the sum of \textit{two} smooth numbers. Here we summarize the main known bounds -- proved and conjectured:

\begin{citedthm}[\cite{balog1989}] Denote $P(x)$ as the largest prime factor of $x$. Then for every sufficiently large integer $N = n_1 + n_2$, it holds that
$$
P(n_1\cdot n_2) \leq N^{\frac{4}{9\sqrt{e}} + \epsilon} =  \mathcal{O}(N^{0.26957}).
$$
\end{citedthm}

A much stronger bound was conjectured by Erd\"os~\cite{erdos1980}:

\begin{conjecture} If $N = n_1+n_2$, then 
\begin{equation}
    P(n_1\cdot n_2) \leq \exp\{c\cdot\sqrt{\log N \log \log N}\},
\end{equation}
where $c$ is an effectively computable constant.
\end{conjecture}

If we consider the representation of sufficiently large integers as the sum of \textit{three} smooth numbers, then the best known upper bound is:

\begin{citedthm}[\cite{sarkozy1995}]
If $N = n_1 + n_2 + n_3$, then 
\begin{equation}
    P(n_1 \cdot n_2 \cdot n_3) \leq  \exp\left\{\sqrt{3/2 + \epsilon}\cdot\sqrt{\log N \log \log N}\right\}.
\end{equation}
\end{citedthm}

It is very interesting that in the case of three smooth integers it is also possible to prove \textit{lower} bounds:

\begin{citedthm}[\cite{sarkozy1995}]
    Let $\epsilon > 0$ be fixed. Then, for every integer $N \geq N_0(\epsilon)$, there exist an integer $n \leq N$, such that every representation of $n$ as $n_1+n_2+n_3$ satisfies the condition
    $$
    P(n_1\cdot n_2 \cdot n_3) \geq (\log N)^{\frac{3}{2}-\epsilon}.
    $$
\end{citedthm}

It is believed (Sárközy) that the truth is much closer to the lower bound, but rigorously proving this will require radically new techniques. Our conjecture is that, in the case of two smooth numbers, the exponent of $\log N$ in the last mentioned theorem is as low as 2. The rigorous proof of this seems considerably more difficult than the proof of Sárközy's bound.

\section{The simplest one-way function}\label{sec:one-way}

\begin{table*}[ht!]
\centering
\caption{Complexity comparison between the most popular one-way functions and the proposed one (in italic).}
\setlength{\tabcolsep}{8pt}
\renewcommand{\arraystretch}{1.3}
\begin{tabular}{m{4.2cm} | m{4,5cm} m{3.6cm} m{3,6cm}} \\[-1ex]
    \textbf{One-way function} & \textbf{Input data} & \textbf{Forward operation} & \textbf{Output complexity} \\ [0.1cm] \hline \\[-1.5ex]
    
    Factoring & Two big primes & One multiplication & $\mathcal{O}(e^{\sqrt[3]{(\log x)(\log\log x)^2}})$\\[3ex]
    
    Discrete Logarithm Problem & One big prime $p$ and a generator of the group $GF(p)$ & One modular exponentiation & $\mathcal{O}(e^{\sqrt[3]{(\log x)(\log\log x)^2}})$\\[3ex]
    
    Elliptic Curve Discrete Logarithm Problem & An elliptic curve over a finite field $GF(p)$ and a point on that curve & One elliptic curve point multiplication & $\mathcal{O}(\sqrt{2^x})$\\[3ex]
    
    \textit{Smooth integers (Problem~\ref{prob:one-way})} & \textit{Two smooth numbers} & \textit{One addition} & $\mathcal{O}\left(e^{\left(\log 4\right)\frac{\log x}{\log\log x}}\right)$\cite{Banks2011}\\[1.5ex]
    \hline\hline
 \end{tabular}
  \label{tab:one-way}

\end{table*}

One-way functions are the cornerstone of the public-key cryptography. For example, the RSA algorithm is based on the conjectured difficulty of factoring problems. The full definition of these one-way function should take into account the complexity of producing two big primes and not only the complexity of their multiplication. Modern computational number theory offers a variety of algorithms for generating big primes, but any of these algorithms requires a large number of multi-word divisions and they are time consuming.\newline

The situation is drastically different, if we work with smooth integers. Whilst it is really easy to produce two large smooth numbers (much easier than, say, to generate two big primes), it appears to be very difficult to solve the following problem:

\begin{problem}[Reversing the sum of two smooth numbers]\label{prob:one-way}
    Given an integer $n$, find a representation of $n$ as the sum of two smooth numbers with the largest prime factor bounded by $\mathcal{O}((\log n)^{2+\epsilon})$, where $\epsilon$ is any positive number.
\end{problem} 

This seems to be the simplest one-way function known so far, since indeed the generation of the input (smooth) numbers and their addition requires practically no efforts at all, whereas solving the reverse problem seems rather difficult. One can use the LLL algorithm~\cite{lenstra1982} to find a smooth number (say, $x$) close to $n$, but the probability that $n-x$ will be also smooth is negligible.\newline

Table~\ref{tab:one-way} summarizes the main characteristics of the most widely used one-way functions and the corresponding complexity figures of merit. It is important to take into account not only the time needed to solve the inverse (hard) problems, but also the time needed to produce the input data and the complexity of the forward (easy) problems.

\section{Generating RSA keys}\label{sec:rsa-keys}
The RSA algorithm requires the generation of two large prime numbers that serve as a secret key for a user. Hence, a fundamental component of RSA key generation is given by primality testing algorithms. In Section~\ref{sec:primality-testing-background}, we provide some relevant background information about primality testing algorithms and their computational complexity. Then, in Section~\ref{sec:primality-generation}, we briefly outline the primality generation problem and our proposal based on the use of smooth integers.

\subsection{Primality testing algorithms}\label{sec:primality-testing-background}
One of the most important problems in computational number theory is the problem on primality testing:
\begin{problem}[Primality testing]
    Given a large integer $p$, determine whether it is a prime or a composite number.
\end{problem}
For large prime numbers, it is clear that the exhaustive search algorithm that tests all the potential prime divisors of $p$ is computationally infeasible. In this section we review the most relevant algorithms used to test primality.\newline

\noindent\textbf{Fermat's Little Theorem.} One can test if
$$
2^{p-1}\equiv 1(\textrm{mod} \ p)
$$
and, if so, then either $p$ is a prime or $p$ is a 2-pseudoprime according to the Fermat’s Little Theorem (FLT). The smallest composite number, for which this test fails is 341. One can substitute 2 with larger values, but still there is a set of composite numbers, called Carmichael numbers, for which the test produces an incorrect answer. The fact that the set of Carmichael numbers is infinite has been established in 1994~\cite{Alford1994}.\newline

\noindent\textbf{Rabin-Miller primality test.} So, instead of using FLT-based tests, we can use more precise Rabin-Miller primality test. If in computing $a^{p-1}(\textrm{mod} \ p)$ one gets ``1'' as an answer, the algorithm performs a ``forensic'' investigation on how exactly this outcome 1 has been obtained. In this case, the one can use only a very small number of witnesses in order to test the primality of $p$, but the proof that only small number of witnesses is sufficient depends on the correctness of the extended Riemann hypothesis.\newline

\noindent\textbf{Solovay-Strassen primality test.} The Solovay-Strassen primality testing algorithm is based on a very simple idea: to test if $p$ is a prime number, one computes $a^{\frac{p-1}{2}}$ and compares this to the value of the Jacobi symbol $\left(\frac{a}{p}\right)$. If $p$ is a prime number, the value of the Jacobi symbol is the same as the value of the Legendre symbol $\left(\frac{a}{p}\right)$. If $p$ is not a prime, then these two values are the same with at most $50\%$ probability. The entire point of the algorithm is that there is no need to factorize $p$ in order to evaluate the Jacobi symbol. So, if the algorithm is executed for, say, 100 values of $a$ and in all the cases
\begin{equation}
    a^{(p-1)/2} \equiv \left(\frac{a}{p}\right)(\textrm{mod} \ p),   
\end{equation}
then we can claim that $p$ is a prime with probability at least $1-2^{-100}$\cite{erdos1986}. The biggest drawback of this algorithm is the necessity to compute the Jacobi symbol, which involves a large number of GCD computations, and is the chief reason why it is rarely used in practice.
\newline

\noindent\textbf{Generalized Fibonacci-based primality test.} A similar algorithm is based on the following interesting property of Fibonacci numbers: \textit{For every prime number, except 5, the following congruence holds:}
$$
F_{p^2-1}\equiv 0(\textrm{mod} \ p).
$$
Since the value of $F_{p^2-1}(\textrm{mod} \ p)$ can be obtained in $\mathcal{O}(\log p)$ operations~\cite{diporto1988, lidl1990}, the algorithm is attractive. Again, it fails for very few, specific composite numbers, called Fibonacci pseudo-primes -- the smallest one being 161.\newline

\subsubsection{Computational complexity of primality testing}

In Table~\ref{tab:primality-test}, we evaluate the computational complexities to test the primality of $p$ for the methods described above. According to our analysis, it is clear that Rabin-Miller's approach is superior:
\begin{itemize}
    \item When comparing Rabin-Miller and Solovay-Strasses tests, we notice that the latter technique requires the \textit{same} number of modular multiplications plus $\ln{p}$ evaluations of the Jacobi symbols, which requires approximately the same computational time.
    \item Fibonacci-based primality testing is implemented by exponentiating the matrix $\begin{pmatrix}1 & 1\\1 & 0\end{pmatrix}$ to the power of $p$. The constant in our estimation, $10.5$, is based on the assumption that one uses Strassen's matrix multiplications algorithm\footnote{The use of standard matrix multiplications algorithm will increase this constant to $12$.}.
\end{itemize}
Therefore, Rabin-Miller is about twice faster than Solovay-Strasses test and about seven times faster than generalized Fibonacci-based primality test. This basically makes Rabin-Miller's test as the de-facto standard in the primality testing field. A similar analysis can be found in the article~\cite{garrett2005}.

\begin{table}[ht!]
\centering
\caption{Complexity to test the primality of $p$ (MM stands for modular multiplications).}
\setlength{\tabcolsep}{8pt}
\renewcommand{\arraystretch}{1.3}
\begin{tabular}{m{2.2cm}|m{5.5cm}}
\textbf{Primality test} & \textbf{Complexity} \\[0.1cm] \hline \\[-1.5ex]
Rabin-Miller        & $1.5\cdot\ln p\cdot\log_2 p$ (MM) \\ [0.1cm] \\[-1.5ex]
Solovay-Strasses   & $1.5\cdot\ln p\cdot\log_2 p$ (MM) + $\ln p$ (Jacobi symbols estimation) \\ [0.1cm] \\[-1.5ex]
Fibonacci-based     & $10.5\cdot\ln p\cdot\log_2 p$ (MM) \\ [0.2cm] \hline\hline
\end{tabular}
\label{tab:primality-test}
\end{table}

For the sake of completeness, we should also mention that the complexities mentioned in Table~\ref{tab:primality-test} are based on the correctness of the extended Riemann hypothesis. In 2004, in their famous article ``PRIMES is in P'', Agrawal et al.~\cite{Agrawal2004} found a primality testing algorithm that works in polynomial time for which computational complexity analysis does \textit{not} depend on any unproved hypothesis. The initial version of the algorithm has a complexity of $\mathcal{O}(\log^{12}p)$, and subsequently improved to $\mathcal{O}(\log^6p)$ after the efforts of many researchers in the field.

\subsection{Primality generation problem}\label{sec:primality-generation}
The following problem is of a fundamental importance in fields like hashing, public-key cryptography, and search of prime factors in large numbers:

\begin{problem}[Generation of large primes]
    Find a large number $p$ which is prime.
\end{problem}

While the primality testing is provably computationally tractable in deterministic polynomial time, for the primality generation we have to assume some strong number-theoretic conjectures to prove the computational efficiency. Even assuming the correctness of the Riemann hypothesis is not enough. But there is conjecture~\cite{cramer1936} that states:

\begin{conjecture}
There is at least one prime in the interval $[x, x+
\ln^2x]$.
\end{conjecture}

Assuming the correctness of this conjecture and the existence of efficient primality testing algorithms, we can easily deduce that the primality generation problem is solvable in polynomial time. 

\subsubsection{Fast primality generation based on the use of smooth integers}\label{sec:smooth-numbers-generation}

The algorithms that are used in the public-key encryption systems to generate prime numbers are based on the following general ideas:

\begin{itemize}
    \item Trial division of the prime number candidate aimed at detecting small prime divisors; the upper limit can be set by the programmer, but in the case of decentralized RSA key generation it is commonly assumed to perform divisions by the first 150 primes and check if none of the divisions produces a residue $0$.
    \item Apply one of the existing probabilistic primality testing algorithms for the numbers that ``survive'' (see Section~\ref{sec:primality-testing-background}).
\end{itemize}

The trial division part is often overlooked from the analysis. However, it is important to point out that the first part of the testing procedure is actually not negligible. In fact, it requires a very large number of multi-word divisions and, depending on the limit of the small primes to be tested, it may take up to 30\% of the actual timing of the entire primality generation procedure (see Section~\ref{sec:computational-complexity}). For instance, the primality testing algorithms used by Ligero MPC protocol~\cite{ames2017} test the first 150 primes (that is, up to 863). More to the point, in the case of decentralized RSA key generation, different nodes are required to produce different numbers and their sum is supposed to be a prime number. If the numbers produced fail to deliver a prime number, the process is repeated.\newline

The properties of the smooth numbers allow us to produce a new algorithm, that can produce RSA keys faster than the existing ones \textit{by removing the need for the trial division part}. In Section~\ref{sec:performance-analysis}, we will provide experimental evidence of the improvements provided by our algorithm against standard primality generation techniques. As mentioned, the idea is to generate smooth numbers to remove the trial division part. Let us consider the set of first 100 primes. Let us also divide this set into two subsets that do not have a common element. For example, $S_1 = {2,5,11,17,23, 
\dots}$ and $S_2 = {3,7,13,19,29,\dots}$. Then we generate a set of small random integer exponents $r_i$, e.g., from the interval $[1,4]$. Now we produce two smooth numbers:
\begin{equation*}
    a=2^{r_1}\cdot 5^{r_3}\cdot 11^{r_5}\dots \quad \text{and} \quad b=3^{r_2}\cdot 7^{r_4}\cdot 13^{r_6}\dots    
\end{equation*}

Then, the sum $a+b$ is for sure not divisible by the first $100$ primes. So, it has considerably higher chances to be a prime in comparison to a randomly selected number of the same size.

\subsubsection{Extension to more than two smooth numbers}

As we have pointed out above, in the case of decentralized RSA key generation there is a need to generalize the solution to more than two smooth numbers. Namely, to produce several (say, a few dozens) random integers, $d_1$, $d_2$, \dots, $d_k$, in a way that their sum will be prime. The above outlined solution for the case $k=2$ (two smooth integers only) can be easily generalized to \emph{any} $k>2$.\newline

Here is one possible solution: let the pair $(d_1, d_2)$ be generated in the way outlined above. For the other numbers, $d_3, d_4, \dots, d_k$, we use smooth numbers of the form:
$$
d_i=2^{h_1}\cdot 3^{h_2}\cdot 5^{h_3}\cdot 7^{h_4}\cdot 11^{h_5}\dots,
$$
where the exponents $h_i$ are randomly chosen integers from the set $[1,2]$. The reason why we use the set $[1,2]$ is that in the selection of $d_1$ and $d_2$ we used only half of the first 100 primes with exponents in $[1,4]$. For the rest of the integers we use all the primes and random exponents 1 or 2; in this case, the sizes of all numbers will be compatible.

\subsubsection{Theoretical bounds}

The theoretical bounds for the number of smooth integers that have to be used are based on the following considerations. Suppose that we would like to generate a $k$-bit prime number (typically, $k = 512, 1024, 2048$). So, we have to find \textit{how many small primes} one needs to multiply in order to obtain a number larger than $2^k$. This can be quantified by using the following Theorem, proved by Sándor and Verroken:

\begin{citedthm}[\cite{sandor2011}]
    Let $s(n)$ denotes the geometric mean of the product of the first $n$-primes. Then
    \begin{equation*}
        \lim \frac{p(n)}{s(n)} = e,
    \end{equation*}
    where $p(n)$ is the $n$-th prime.
\end{citedthm}

This theorem allows us to obtain a closed formula for the expected number of smooth integers that have to be used as a function of the size of the key-to-be-generated: if we have to generate a $k$-bit RSA key, then it is sufficient to use the \textit{first} $k/\ln k$ small primes to produce the smooth numbers in our algorithm.\newline

In Table~\ref{tab:small-primes}, we compare the theoretical bounds and the exact results from the numerical simulations. One can immediately make the following conclusions:

\begin{itemize}
    \item The number of small primes required by our smooth numbers techniques is slightly smaller than the number predicted by the theory.
    \item The \textit{difference} can be explained if one follows very carefully the proof of the theorem from the paper above.
\end{itemize}

Our understanding is that for a practical application of our algorithm, the users of this technique do not have to become experts on analytic number theory. The main point of the article is to provide the algorithm in as understandable as possible format and, for more mathematically oriented readers, we provide more than enough references to get deeper into this fascinating issue.  

\begin{table}[ht!]
\centering
\caption{Theoretical and experimental study of the number small primes needed depending on key size.}
\setlength{\tabcolsep}{8pt}
\renewcommand{\arraystretch}{1.3}
\begin{tabular}{m{4.8cm}|>{\centering}m{0.6cm}>{\centering}m{0.6cm}>{\centering\arraybackslash}m{0.6cm}}
        \\[-1ex]
\textbf{Key size (in bits)}                             & \textbf{512}  & \textbf{1024} & \textbf{2048}  \\[0.1cm] \hline \\[-1.5ex]
Theoretical prediction for the number of small primes needed    & 82            & 148           & 269   \\ \\[-1.5ex]
Exact bounds based on computational experiments                 & 75            & 135           & 239   \\[0.3cm] \hline\hline
\end{tabular}
\label{tab:small-primes}
\end{table}

\section{Experimental performance analysis}\label{sec:performance-analysis}

In this section, we validate the benefits of our algorithm compared to the state-of-the-art techniques in primality generation. In Section~\ref{sec:prime-probability}, we analyse the probability of a smooth integer to be prime, and we compare it with a random number; then, in Section~\ref{sec:computational-complexity}, we study the actual savings of the primality generation procedure; finally, in Section~\ref{sec:discussion-results}, we provide a discussion on how to interpret our experimental results, highlighting the fact that our algorithm \textit{always} provides improvements (some times more, some times less) compared to standard techniques.

\subsection{Probability of generating a prime}\label{sec:prime-probability}

We used the first 148 primes for our test and the parameters provided above to produce one million pairs of smooth numbers, $a$ and $b$, and test the primality of their sum. The size of the numbers produced is, on average, $1024$ bits as shown by Table~\ref{tab:small-primes}.\newline

The prime number theorem says that the density of primes less than a given bound $x$ is $\frac{x}{\ln x}$. So, if $x$ is $2^{1024}$, the probability that a randomly selected number less than $x$ would be prime is about $1$ in $710$. Since we will test only odd numbers, obviously we have a chance about $1$ in $355$ ($0.28$\%) to find a prime\footnote{Primality testing algorithms normally requires up to a few hundred microseconds, it is reasonable to expect that in a sub-second period, we will be able to obtain a prime number of this size.}.\newline

On the other hand, the numbers produced as the sum of two smooth numbers have a chance for primality around $2\%$. In order to estimate the probability that an integer less than $x$ is free from prime factors larger than $y$, we can use theorems from de Bruijn~\cite{debruijn1951} or Hildebrand~\cite{hildebrand1986}\footnote{Interestingly, the two papers have exactly the same title.}. Based on those, the probability that a randomly chosen integer less than $x$ does not have prime factors larger than $y$ is given by the formula:

\begin{equation*}
    \mathbb{P}[ n \leq x, P(n) < y] = \frac{\exp(-u/2)}{\log y},
\end{equation*}
where $u = \log x /\log y$  and $P(n)$ denotes the largest prime factor of $n$.  In our case $x$ can be taken as $2^{1024}$ and $y$ is the limit we impose on the largest prime factor in the trial division part of the algorithm. If we take the same estimates as in the case of typical decentralized RSA key generation algorithms (e.g., 1024-bit primes, therefore, $x=2^{1024}$, $y = 853$), then we obtain the bounds mentioned above~\cite{boneh1997}.

\begin{table*}[ht!]
    \centering
    \tabcolsep=0.11cm
    \caption{Experimental data based on the generation of two co-prime smooth 100-integers.}
    \begin{tabular}
        {m{7,5cm}|>{\centering}m{1,4cm}>{\centering}m{1,4cm}>{\centering}m{1,4cm}>{\centering\arraybackslash}m{1,4cm}}
        \\[-1.5ex]
        \textbf{Exponent range} & $(1,2)$ & $(1,2,3)$ & $(1,2,3,4)$ & $(1,2,3,4,5)$\\[0.2cm] \hline \\[-1.5ex]
        Average size of the numbers generated & 450 bit & 712 bit & 1000 bit & 1313 bit\\[0.2cm] \\[-1.5ex]
        Probability to obtain a prime number generated by adding two 100-integers & 2.9\% & 2.1\% & 1.7\% & 1.45\%\\[0.3cm] \hline\hline
    \end{tabular}
    \label{tab:1}
\end{table*}

\begin{table*}[ht!]
    \centering
    \tabcolsep=0.11cm
    \caption{Expected savings in terms of number of modular exponentiations through the usage of smooth numbers.}
    \begin{tabular}
        {m{7,5cm}|>{\centering}m{1,4cm}>{\centering}m{1,4cm}>{\centering}m{1,4cm}>{\centering\arraybackslash}m{1,4cm}}
        \\[-1.5ex]
        \textbf{Exponent range} & $(1,2)$ & $(1,2,3)$ & $(1,2,3,4)$ & $(1,2,3,4,5)$\\[0.2cm] \hline \\[-1.5ex]
        $(a)$ Expected number of modular exponentiations to test a random integer for primality & 156.5 & 243.9 & 344.8 & 476.2\\[0.45cm] \\[-1.5ex]
        $(b)$ Expected number of modular exponentiations to test for primality an integer produced by our algorithm  & 34.5 & 47.6 & 58.8 & 69.0\\[0.45cm] \\[-1.5ex]
        Improvement factor $(a)/(b)$ & 4.53 & 5.13 & 5.86 & 6.90\\[0.2cm]
        \hline\hline
    \end{tabular}
    \label{tab:2}
\end{table*}

\subsection{Savings in the number of modular exponentiations}\label{sec:computational-complexity}

The trial division part tests if the number is divisible by a small prime up to a given level and, after that, one applies a primality testing algorithm. Rabin-Miller, Solovay-Strassen, Fermat and Fibonacci-based tests all use modular exponentiations. Hence, it is essential to know – as much accurately as possible – the time ratio between one modular exponentiation and one division by a small prime to evaluate the expected savings provided by our algorithm. We highlight here that this is platform-dependent and also depends on how well optimized divisions by a small constant are. For numbers of size $512$-bit we have experimented (Mathematica\footnote{Wolfram Mathematica (\url{https://wolfram.com/mathematica/}).}, Python and C++ with GMP) and have found a ratio about $40:1$, whereas for $1024$-bit numbers the ratio is about $1000:1$.\newline

Table~\ref{tab:1} provides some experimental data, obtained as follows. We produce two co-prime smooth $100$-integers, that is, their largest prime factor is the $100th$ prime and also impose restrictions on the exponent used. The table provides the interval for the exponents, the average size of the numbers produced and – the most important component – the probability that the sum of these two numbers will be prime. Table~\ref{tab:2} showcases the expected savings in terms of the number of modular exponentiations, needed to be performed – on average – until finding a prime number. The sizes of the numbers are like in the previous table.\newline

\subsubsection{Detailed computational complexity analysis}

Existing RSA key generation algorithm contains trial divisions to test the existence of small prime factors, and a primality testing algorithm, which takes exactly one modular exponentiation. Hence, the computational complexity of such techniques, which we denote by $cc$, is based on the formula:
\begin{equation*}
    cc = td + me,
\end{equation*}
where $td$ is the number of trial divisions and $me$ is the number of modular exponentiations.\newline

Our proposal to fast generate primes uses the same number of modular exponentiations but replaces the trial divisions with the generation of random numbers based on smooth integers. That involve only multiplications between pre-computed powers of small primes. So, the computational complexity for our algorithm, denoted by $cc^*$, is given by:
\begin{equation*}
    cc^* = me + sm,
\end{equation*}
where $sm$ is the number of multiplications needed for the generation of smooth integers.\newline

Let us calculate the computational complexities in the case of 1024-bit RSA private keys, which, of course, need to generate a 1024-bit prime number. First, we compute the number of trial divisions $td$. With our technique, we produce a random number with a chance for being prime about 1 out of 50. Hence, we need on average $50$ numbers to survive the trial divisions before succeeding the Rabin-Miller's test. That means that we have computed all the 150 trial divisions by small primes 50 times, i.e., $150\cdot 50 = 7500$ trial divisions. Furthermore, since the probability that a randomly chosen odd number of this size is prime is about 1 out of 350, we have to consider that we will perform only one division when we test divisibility by $3$ (i.e., $1/3$ of the time), only two divisions when the number is divisible by $5$ but not by $3$, and so on. It is possible to calculate that the average number of trial divisions is actually only $5$. So we need extra $300\cdot 5 = 1500$ divisions. To sum up, the number of modular exponentiations for the two approaches is the same, but:
\begin{itemize}
    \item With trial division based primality testing algorithms one needs approximately 9000 (\textit{expensive}) multi-word divisions;
    \item With our algorithm one needs approximately 7500 (\textit{inexpensive}) multiplications.
\end{itemize}

Table~\ref{tab:improvement-factor} showcases the time savings when using our algorithm compared to a trial-division based algorithm. The library used is NTL\footnote{NTL: A Library for doing Number Theory (\url{https://libntl.org/}).}. The results with Mathematica are almost the same. Again, if one uses different libraries, the improvements can be quite different, but for sure there will be \textit{improvements}.\newline

\subsection{Discussion of the results}\label{sec:discussion-results}

The exact savings depend on the ratio between modular exponentiations timings and multi-word division timings. In sharp contrast with the case of modular inversion (over prime fields)  and modular multiplication, when the ratio 80:1 is usually assumed as a standard, in this case we have no universally accepted timing ratio for these two problems: modular exponentiations and multi-word divisions. And it is clear that this ratio depends on the dynamic range of the computations. For example, the decentralized key generation part of the Ligero MPC protocol~\cite{ames2017} uses primes up to $853$. This is exactly the $150th$ prime number. In other computational number theory products it is recommended to test primes up to $100$, which takes only six times less for this particular part of the algorithm. More to the point, if one attempts to generate a $1024$-bit prime, the time for the modular exponentiations (necessary to execute the Rabin-Miller test) is considerably smaller (about eight times) in comparison to the $2048$-bit modular exponentiation. In the latter case, the contribution of the trial division part of the primality testing algorithm will be much smaller in comparison to the former case. To summarize, in the case of modular exponentiations, doubling the size of the exponent and the modulo leads to increasing the modular exponentiation timing by a factor of about eight, whereas for the trial division the multiplicative incremental factor is only about two. This is the reason why our algorithm saves a lot in the case of 512-bit primes, less in the case of 1024-bit primes and only marginally in the case of 2048-bit primes. The timings that we have reported are based on our experiments with NTL. It is clear that with different libraries one can get different numbers. So, the important facts -- from a practical point of view -- are:

\begin{itemize}
\item The algorithm \textit{always} leads to savings in comparison to algorithms that use trial divisions.
\item The savings become smaller as the size of the primes-to-be-generated becomes larger;
\item the actual savings can greatly differ depending on what library is used for the actual implementation.
\item The \textit{most} critical component is the ratio: time for a modular exponentiations over time for a multi-word division. 
\end{itemize}

\begin{table}[ht!]
\centering
\caption{Time savings when using our algorithm for different key sizes.}
\setlength{\tabcolsep}{10pt}
\renewcommand{\arraystretch}{1.3}
\begin{tabular}{l|cccc} \\ [-1ex]
\textbf{Key size (in bits)} & \textbf{512}  & \textbf{1024} & \textbf{2048} & \textbf{4096}  \\ [0.1cm] \hline \\[-1.8ex]
Time savings                  & 30.4\%        & 12.1\%        & 2.9\%         & 0.8\% \\ [0.1cm] \hline\hline
\end{tabular}
\label{tab:improvement-factor}
\end{table}

\section{Real-world applications}\label{sec:practical-considerations}

\subsection{Implementation}

The RSA key generation algorithm proposed in the paper is rather straightforward to implement. To summarize, one generates several smooth numbers based on the following conditions: two of the smooth numbers are co-prime, the other numbers are smooth and divisible by all small prime numbers up to the limit selected and all these numbers are added. After that, one directly applies the Rabin-Miller testing, since it is guaranteed that there is no need for trial divisions. A programmer interested in using the technique has to do only one thing, that is to check the time ratio between the trial multi-word divisions and the time for generating and adding smooth numbers. Since those procedures will differ vastly over different platforms, the only pragmatic judge of the computational savings is by direct computational experiments.\newline

Programmers have realized the necessity to avoid trial divisions, if possible, some time ago and there were other attempts to achieve this goal. The most (in)famous example is, perhaps, the primality generation software at the RSALib. This library is widely used in practice. Until just a few years ago, they used the following procedure to produce random numbers that are subject to Rabin-Miller's test. The number $p$ is generated as
\begin{equation}
    p = k \cdot M + (65537^a \bmod M),
\end{equation}
where $M$ is the product of the first $n$ small primes. $n=39$ is used to generate primes with binary length $[512,960]$ bits, while $n = 71$, $126$ and $225$ are used to generate random numbers with binary lengths $[992,1952]$, $[1984,3936]$ and $[3964,4096]$ respectively. The two parameters, $k$ and $a$ are unknown and randomly selected.\newline

This way of producing random numbers guarantees that the numbers produced will be NOT divisible by the first $n$ primes. However, it has been discovered~\cite{nemec2017} that this algorithm for producing random numbers generates keys that can be successfully revealed by the Coppersmith's attack~\cite{coppersmith1997}. The reason is that this method for key generation leads to an unintended disclosure of sufficiently many bits of the secret keys that allows a successful application of this attack. After the discovery of that fatal flaw, the key generation procedures in RSALib were immediately modified. The reader can find all the details in~\cite{nemec2017}.\newline

The algorithm proposed in our paper does not reveal any bits of the keys and produces random numbers with very high entropy. Thanks to that it is secured against the Coppersmith's attack, which is another positive characteristic of it. 

\subsection{Application scenario}

There are at least three practical scenarios that can benefit from the findings in this paper.

\subsubsection{Classical implementation of RSA algorithm}
The selection of the main parameters of the RSA algorithm -- the private and public keys -- can be done faster if one uses the techniques proposed by us. The idea to use shared RSA keys was proposed first by Boneh and Franklin 24 years ago in~\cite{boneh1997}. This is the first paper that actually considers the problem: how to generate a certain amount of random numbers (produced by different users privately and independently) in such a way as to maximize the chance that their sum would be a prime number. Since the specifics of the Boneh-Franklin protocol requires the primes to be of the form $4k+3$, the initial solution that they propose is this: one user produces a random number congruent to $3 (\textrm{mod} \ 4)$, the other users produce random numbers divisible by 4 and they implement secure multi-party summation algorithm afterwards. In this case, if we need 1024-bit shared RSA keys, we will need approximately 350 communications rounds in order to have a reasonable chance to obtain a secret RSA key (e.g., a 1024-bit prime). If the users use our smooth number technique, then the number of rounds will be reduced to only about 50. This is the reason why we claim the reduction of the communication complexity.\newline

However, we admit that the original paper by Boneh and Franklin was aimed at showcasing the principle possibility of sharing RSA keys in a secure manner and not on the implementation specifics. It is the implementation specifics (actually, only some of them) that we address. Since the goal of the original paper was different from ours, we do not include here a direct comparison. The previous analysis clearly outlines the advantages of the new proposed technique.

\subsubsection{Decentralized environments}
Our approach is particularly attractive for decentralized environments. In this case we offer a fast procedure for different parties to produce shared RSA keys. In this case the reduction of the communication complexity is considerably more important than the minor computational complexity savings. This is a very essential problem in modern blockchain architectures, where different parties may want to have (and use) the \textit{same} public key, but this key has to be generated in a way that does not allow any of the parties to have access to the private keys. If different parties make use of the technique based on smooth integers that we propose in the article, this generation can be accomplished in a rather efficient manner.

\subsubsection{Verifiable delay functions}Since the introduction of the so-called Verifiable Delay Functions (VDFs) in 2018~\cite{boneh2018}, there has been a significant interest in this cryptographic primitive. These functions can be used in many applications aimed at creating protections against denial of service attacks or generating random numbers in a distributed way, to name but a few. One of the main showstoppers (perhaps, the only showstopper) for the successful applications of VDFs is the slow generation of shared keys. Our algorithm is a step forward in removing this obstacle.\newline

We have to admit that in order to make VDFs widely used, we need more speed ups than the one we offer here; still, our approach, combined with additional computational optimizations, could lead to an ultimate solution of this problem. We have done research in this field over the last few months and published one article that clarifies specific details~\cite{attias2020}.

\section{Open problems}\label{sec:open-problems}
In this paper we have showcased one possible application of smooth integers, aimed at a faster generation of large prime numbers. The most direct application is the generation of the RSA keys, but of course any other cryptographic schemes that require a fast generation of prime numbers can benefit from such a discovery. We have also provided a lot better estimate of the number of iterations sufficient to find a sparse representation of a given number as the sum of s-integers.\newline
 
It is our understanding that the most important part of the article is the proposition of the new one-way function. From a purely arithmetic point of view, it has a much simpler computational description, since one only needs to generate two smooth numbers and add them. There are at least two potential applications of such a function:
 
\begin{itemize}
    \item One of the most powerful computational complexity features of the one-way function, discovered in~\cite{haastad1999pseudorandom}, is that any one-way function can be used to build an efficient pseudo-random number generator (PRNG). The exact, optimized design of such a PRNG is a matter of on-going research within our group.
    \item Since our one-way function is dual to the factoring problem (e.g., instead of prime numbers we use numbers with small prime factors only, and instead of multiplication we use addition), it seems natural to try to develop a new, competitive to RSA public-key encryption system. Whilst there is no theoretical result that guarantees the existence of such an encryption system, the extreme simplicity of the proposed one-way function is an appealing feature. We hope that this will inspire researchers in the field of public-key cryptography, and computational number theory in attempting to discover such an encryption scheme.
\end{itemize}

\section{Conclusions}\label{sec:conclusions}

In this paper we showcase the possible use of smooth integers for various cryptographic problems of substantial importance in the world of decentralized ledgers.\newline
 
On a concrete level, we propose an algorithm for selecting shares of different communicating parties in a way that can significantly reduce the communication and computational cost of producing RSA keys. With the increase of the dynamic range, the savings provided by our algorithm decrease. But it is important to clarify that it does not slow the performance of the primality testing procedures, simply the elimination of the trial division phase is becoming less and less important as the size of the numbers to be tested for primality increase. This is due to the high computational cost of the modular exponentiations compared to the division by small constants.\newline

On a more abstract level, we showcase an unusually simple one-way function to be researched either as a tool for creating a new public-key cryptosystem, as a hash function, as a cryptographic puzzle. Or maybe something else?

\appendix
The example proposed in Table~\ref{tab:appendixa} is a representation of one of the factors of the RSA challenging number RSA-$512$. This is a $65$-decimal digit prime number ($p_{65}$) used to just showcase its representation as the sum of $5$-integers. This example demonstrates in a pictorial way the work of the greedy algorithm described in the sketch of proof of Theorem~\ref{th:representation} aimed at finding short smooth integer representations. After every single iteration of the greedy algorithm, the number of bits of the difference between the original number and the closest to it smooth integer is reduced in accordance to Theorem~\ref{th:tijdeman74}.\newline

\begin{table*}[ht!]
\centering
\caption{65-decimal digit prime number as a sum of 5-integers.}
\begin{tabular}{r|l}
    \multicolumn{1}{c}{\textbf{Decimal representation}} & \multicolumn{1}{|c}{\textbf{Smooth integer representation}}\\ [0.5ex] \hline \\ [-1ex]
    $\seqsplit{32769 \ 13299 \ 32667 \ 09549 \ 96198 \ 81908 \ 34461 \ 41317 \ 76429 \ 67992 \ 94253 \ 97982 \ 88533} \ =$ & $p_{65} \ =$\\ [2ex]
    $\seqsplit{32769 \ 08112 \ 87987 \ 68937 \ 24041 \ 66538 \ 86611 \ 18749 \ 39274 \ 79410 \ 68800 \ 00000 \ 00000} \ +$ & $2^{35} \cdot3^5 \cdot 5^{12} \cdot7^{27} \cdot 11^{19} \ +$\\ [1ex]
    $\seqsplit{5186 \ 44385 \ 71792 \ 61074 \ 82107 \ 95031 \ 40507 \ 69563 \ 67538 \ 85534 \ 61760 \ 00000} \ +$ & $2^{65} \cdot 3^{49} \cdot 5^{6} \cdot 7^{10} \cdot 11^{3} \ +$\\ [1ex]
    $\seqsplit{293 \ 68803 \ 30536 \ 12792 \ 81357 \ 20414 \ 02265 \ 00362 \ 24000 \ 00000 \ 00000} \ +$ & $2^{54} \cdot 3^{12} \cdot 5^{13} \cdot 7^{13} \cdot 11^{10} \ + $\\ [1ex]
    $\seqsplit{16 \ 80545 \ 49409 \ 03596 \ 65647 \ 74917 \ 97216 \ 46080 \ 00000 \ 00000} \ +$ & $2^{59} \cdot 3^1 \cdot 5^{11} \cdot 7^8 \cdot 11^{13} \ +$\\ [1ex]
    $\seqsplit{71059 \  57613 \ 12070 \ 84069 \ 84528 \ 86917 \ 41015 \ 48800} \ +$ & $2^{20} \cdot 3^5 \cdot 5^2 \cdot 7^5 \cdot 11^3 \ +$\\ [1ex]
    $\seqsplit{2673 \ 33024 \ 83120 \ 63414 \ 38977 \ 63840} \ +$ & $ 2^{14} \cdot 3^{13} \cdot 5^1 \cdot 7^2 \cdot 11^{15} \ +$\\ [1ex]
    $\seqsplit{3381 \ 10622 \ 71257 \ 88075 \ 62240} \ + $ & $ 2^{17} \cdot 3^{19} \cdot 5^1 \cdot 7^9 \cdot 11^1 \ +$\\ [1ex]
    $\seqsplit{5322 \ 84032 \ 30911 \ 85664} \ + $ & $ 2^{14} \cdot 3^9 \cdot 5^0 \cdot 7^5 \cdot 11^5 \ +$\\ [1ex]
    $\seqsplit{14207 \ 36868 \ 32616} \ + $ & $ 2^3 \cdot 3^8 \cdot 5^0 \cdot 7^5 \cdot 11^5 +$\\ [1ex]
    $\seqsplit{75139 \ 85456} \ + $ & $ 2^6 \cdot 3^6 \cdot 5^0 \cdot 7^0 \cdot 11^5 \ +$\\ [1ex]
    $\seqsplit{8 \ 45152} \ + $ & $ 2^5 \cdot 3^0 \cdot 5^0 \cdot 7^4 \cdot 11^1 \ +$\\ [1ex]
    $1155 \ +$ & $ 2^0 \cdot 3^1 \cdot 5^1 \cdot 7^1 \cdot 11^1 \ +$\\ [1ex]
    $10 \quad$ & $2^1 \cdot 3^0 \cdot 5^1 \cdot 7^0 \cdot 11^0$\\ [0.1ex]
 \end{tabular}
\label{tab:appendixa}
\end{table*}

We offer some brief numerical data with the same testing number ($p_{65}$) in the case $s=2,3,4,5$ and compare the findings with the conjecture that the complexity of the greedy
algorithm aimed at finding the representation of a number $n$ as the sum of $s$-integer terminates after about $(\frac{2}{s}+o(1))\frac{\log n}{\log\log n}$ steps. We applied the same program used to find representations as the sum of 5-integers, in Table~\ref{tab:operations_cost}.

\begin{table}[ht!]
\centering
\caption{Comparison between the actual and the predicted number of s-integers produced by the greedy algorithm to represent the test number $p_{65}$.}
\setlength{\tabcolsep}{8pt}
\renewcommand{\arraystretch}{1.3}
\begin{tabular}{m{4.1cm}|>{\centering}m{0.5cm}>{\centering}m{0.5cm}>{\centering}m{0.5cm}>{\centering\arraybackslash}m{0.5cm}}
        \\[-1ex]
\textbf{s-integers}                                                         & \textbf{2}    & \textbf{3}    & \textbf{4}    & \textbf{5}  \\[0.1cm] \hline \\[-1.5ex]
$s$-integers obtained by greedy algorithm                                   & 29            & 20            & 16            & 13 \\ \\[-1.5ex]
Theoretical prediction of number of terms $\frac{2\log n}{s\log\log n}$     & 26.68         & 18.45         & 13.84         & 11.07 \\[0.3cm] \hline\hline
\end{tabular}
\label{tab:operations_cost}
\end{table}

\section*{Acknowledgements}
We thank Prof. Bill Buchanan from Edinburgh Napier University  and Prof. Igor Shparlinski from The University of New South Wales, Australia,
for their comments that greatly improved the manuscript.

\nocite{*}
\bibliographystyle{IEEEtran}
\bibliography{references}

\end{document}